\documentclass[runningheads]{llncs}
\usepackage[german,american]{babel}
\usepackage[T1]{fontenc}
\usepackage[latin1]{inputenc}
\usepackage[T1]{xcolor}
\usepackage{booktabs}
\usepackage{multirow}
\usepackage{amsmath}
\usepackage{amssymb}

\usepackage{graphicx}
\DeclareGraphicsExtensions{.pdf,.ai,.jpg,.png}
\setkeys{Gin}{pagebox=artbox}
\graphicspath{{.}}

\newcommand{\bsfigure}[3][scale=1.0]{%
  \begin{figure}[tb]
    \centering
    \includegraphics[#1]{#2}
    \vspace{-1ex}
    \caption{#3}\label{#2}
    \vspace{-3ex}
  \end{figure}}

\newcommand{\query}[1]{\texttt{\small #1}}

\newcommand{\Ni}{(1)~}
\newcommand{\Nii}{(2)~}
\newcommand{\Niii}{(3)~}

\newcommand{\Na}{(a)~}
\newcommand{\Nb}{(b)~}

\usepackage[numbers,sectionbib]{natbib} 
\begin{document}

\title{\mbox{\kern-.5em How Train--Test Leakage Affects Zero-shot Retrieval}}
\titlerunning{How Train--Test Leakage Affects Zero-shot Retrieval}

\author{Maik Fr{\"o}be,\inst{1} Christopher Akiki,\inst{2} Martin Potthast,\inst{2} Matthias Hagen\inst{1}}
\authorrunning{Fr{\"o}be et al.}
\institute{Martin-Luther-Universit{\"a}t Halle-Wittenberg \and Leipzig University}

\maketitle
\begin{abstract}
Neural retrieval models are often trained on (subsets of) the millions of queries of the MS~MARCO / ORCAS datasets and then tested on the 250~Robust04 queries or other TREC~benchmarks with often only 50~queries. In such setups, many of the few test queries can be very similar to queries from the huge training data---in fact, 69\% of the Robust04~queries have near-duplicates in MS~MARCO~/~ORCAS. We investigate the impact of this unintended train--test leakage by training neural retrieval models on combinations of a fixed number of MS~MARCO~/~OR\-CAS queries that are highly similar to the actual test queries and an increasing number of other queries. We find that leakage can improve effectiveness and even change the ranking of systems. However, these effects diminish the smaller and, thus, the more realistic the amount of leakage is among all training instances.

\keywords{Neural information retrieval; Train--test leakage; BERT; T5}
\end{abstract}

\section{Introduction}

Training transformer-based retrieval models requires large amounts of data unavailable in many traditional retrieval benchmarks~\cite{mokrii:2021}. Data-hungry training regimes became possible with the 2019 release of MS~MARCO~\cite{craswell:2019} and its 367,013~queries that were subsequently enriched by the ORCAS click log~\cite{craswell:2020} with 10~million queries. Fine-tuning models trained on MS~MARCO to other benchmarks or using them without fine-tuning in zero-shot scenarios is often very effective~\cite{mokrii:2021,nogueira:2020,zhang:2020}. For example, monoT5~\cite{nogueira:2020}, which has been trained only on MS~MARCO data, is currently the most effective model for the Robust04 document ranking task.%
\footnote{\url{https://paperswithcode.com/sota/ad-hoc-information-retrieval-on-trec-robust04}}
Furthermore, the reference implementations of monoT5 and \mbox{monoBERT}~\cite{nogueira:2019} in retrieval frameworks such as PyTerrier~\cite{macdonald:2021} or Pyserini / Py\-Gaggle~\cite{lin:2021b} all use models trained only on MS~MARCO by default. However, when MS~MARCO was officially split into train and test data, cross-benchmark use was not anticipated, so that MS~MARCO's training queries may overlap with the test queries of other much smaller datasets (e.g., Robust04). In this paper, we investigate the impact of such a train--test leakage by training neural models on MS~MARCO document ranking data with different proportions of controlled leakage to Robust04 and the TREC~2017 and 2018~Common Core tracks as test datasets.

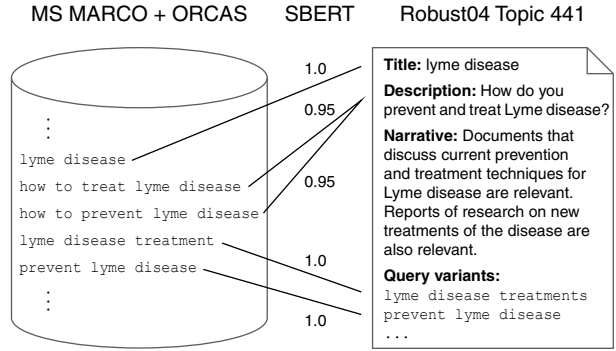
\bsfigure{example-leaking-queries-ms-marco-robust04-topic-441}{MS~MARCO / ORCAS queries with high Sentence-BERT~(SBERT) similarity to Robust04 Topic~441.}

To identify probably leaking queries, we run a semantic nearest-neighbor search using Sentence-BERT~\cite{reimers:2019} and compare each MS~MARCO / ORCAS query to the title, description, and manual query variants~\cite{benham:2017,benham:2018} of the topics in Robust04 and the TREC~2017 and 2018 Common Core tracks. Figure~\ref{example-leaking-queries-ms-marco-robust04-topic-441} illustrates this procedure for Topic~441 (\query{lyme\;disease}) from Robust04. Our manual review of the leakage candidates shows that 69\%~to~76\% of the topics have near-duplicates in MS~MARCO / ORCAS. To analyze the effect of this potential train--test leakage on neural retrieval models, we create three types of training datasets per test corpus, in variants with~1,000 to 128,000~training instances (query~+~(non-)relevant document):
\Ni
a fixed number of instances derived from test queries from the test corpora (1000~for Robust04 and 200~for each of the two Common~Core tracks), augmented by other random non-leaking MS~MARCO / ORCAS instances to simulate an upper bound on train--test leakage effects,
\Nii
a fixed number of leaking MS~MARCO~/~OR\-CAS instances (1000~for Robust04 and 200~each for the two Common Core tracks) supplemented by other random non-leaking instances, and
\Niii
random MS~MARCO / ORCAS instances, ensuring that no train--test leakage candidates are included.

In our experiments, we observe leakage-induced improvements in effectiveness for Robust04 and the two Common Core tracks, which can even change the ranking of systems. However, the average improvements in overall effectiveness are often not significant and decrease as the proportion of leakage in the training data becomes smaller and more representative of realistic training scenarios. Nonetheless, our experiments on the effects of leaked instances on search results and the resulting system rankings show that leakage effects occur even when improvements in effectiveness are statistically negligible---a strong argument that train--test leaks should be avoided in academic experiments.%
\footnote{All code and data is publicly available at \url{https://github.com/webis-de/SPIRE-22}.}

\section{Background and Related Work}

Disjoint training, validation, and test datasets are essential to properly evaluate the effectiveness of machine learning models~\cite{chollet:2021}. Duplication between training and test data can lead to incorrectly high ``effectiveness'' by memorizing instances rather than learning the target concept. In practice, though, train and test data often still contain redundancies. For text data, paraphrases, synonyms, etc., can be especially problematic, resulting in train--test leaks~\cite{he:2009,krishna:2021,linjordet:2020}. For instance, the training and test sets of the ELI5~dataset~\cite{fan:2019} for question answering were created using TF-IDF as a heuristic to eliminate redundancies between them. This proved insufficient as 81\%~of the test questions turned out to be paraphrases of training questions, which clearly favored models that memorized the training data~\cite{krishna:2021}. Recently, \citet{zhan:2022} found that 79\%~of the TREC 2019~Deep Learning track topics have similar or duplicated queries in the training data and proposed new data splits to evaluate the interpolation and extrapolation effectiveness of models. However, not all types of train--test leaks are unintentional. The TREC~2017 and~2018 Common Core tracks~\cite{allan:2017} intentionally reused topics from Robust04 to allow participants to use the relevance judgments for training. Indeed, approaches trained on the Robust04~judgments were more effective than others~\cite{allan:2017}. In this paper, we study whether a similar effect can be observed for unintentional leakage from the large MS~MARCO and ORCAS~datasets.

Training retrieval models on MS~MARCO and applying them to another corpus is a form of transfer learning~\cite{mokrii:2021}. Transfer learning is susceptible to train--test leakage since the train and test data are often generated independently without precautions to prevent leaks~\cite{chen:2020}. Research on leakage in transfer learning focuses on membership inference~\cite{nasr:2019,shokri:2017} (predicting if a model has seen an instance during training) and property inference~\cite{ateniese:2015,fredrikson:2015} (predicting properties of the training data). Both inferences rely on the observation that neural models may memorize some training instances to generalize through interpolation~\cite{berthelot:2019,chollet:2021} and to similar test instances~\cite{feldman:2020a,feldman:2020b}. It is unclear whether and how neural retrieval models in a transfer learning scenario are affected by leakage. Memorized relevant instances might reduce effectiveness for different test queries while improving it for similar queries, like the examples in Figure~\ref{example-leaking-queries-ms-marco-robust04-topic-441}. We take the first steps to investigate the effects of such a train--test leakage.

When the target corpus contains only few training instances, transferred retrieval models are often more effective without fine-tuning, in a zero-shot setting~\cite{zhang:2020}; for instance, when training on MS~MARCO and testing on TREC datasets~\cite{mokrii:2021,nogueira:2020,zhang:2020}. A frequently used target TREC~dataset is Robust04~\cite{voorhees:2005} with 250~topics and a collection of 528,155~documents published between 1989 and~1996 by the Financial Times, the Federal Register, the Foreign Broadcast Information Service, and the LA~Times.%
\footnote{https://trec.nist.gov/data/cd45/index.html}
Later, the TREC Common Core track~2017~\cite{allan:2017} reused 50~of the 250~Robust04~topics on the New York Times Annotated Corpus~\cite{sandhaus:2008}%
\footnote{\url{https://catalog.ldc.upenn.edu/LDC2008T19}}
(1,864,661~documents published between~1987 and~2007) and the Common~Core track~2018 reused another 25~Robust04 topics (and introduced 25~new topics) on the Washington Post Corpus%
\footnote{\url{https://trec.nist.gov/data/wapost/}}
(595,037~documents published between 2012~and~2017). A total of 311,410~relevance judgments were collected for the Robust04 topics, 30,030~for the TREC~2017 Common Core track, and~26,233 for the TREC~2018 Common Core track. Interestingly, every Robust04~topic and every topic from the Common Core tracks~2017 and~2018 was augmented with at least eight query variants compiled by expert searchers, and made available as an additional resource~\cite{benham:2017,benham:2018}.

Research on paraphrase detection~\cite{dolan:2005,wahle:2021} and semantic question matching~\cite{sharma:2019} is of great relevance to the identification of potentially leaking queries between training and test data. \citet{reimers:2019} and \citet{lin:2020} showed that pooling or averaging the output of contextual word embeddings of pre-trained transformer encoders like BERT \cite{devlin:2019} is not suited for paraphrase detection---both, with respect to efficiency and accuracy. Sentence-BERT~\cite{reimers:2019} solves this issue by adopting a BERT-based triplet network structure and a contrastive loss that attempts to learn a global and a local structure suited for detecting semantically related sentences. We therefore use Sentence-BERT in a version specifically trained for paraphrase detection to identify leaking queries.

\section{Identifying Leaking Queries}

To examine the impact of possible leaks from MS~MARCO / ORCAS to the TREC~datasets Robust04 and Common Core~2017 and~2018, we compare the former's queries (367,013 plus 10~million) to the 275~topics of the latter three. Since lexical similarity may not be sufficient, as indicated by the ELI5 issue~\cite{krishna:2021} mentioned above, we compute semantic similarity scores using Sentence-BERT~\cite{reimers:2019}.%
\footnote{Of the available pre-trained Sentence-BERT models, we use the paraphrase detection model: \url{https://huggingface.co/sentence-transformers/paraphrase-MiniLM-L6-v2}}
We store the Sentence-BERT embeddings of all MS~MARCO and ORCAS queries in two Faiss indexes~\cite{johnson:2021} and query them for the 100~nearest neighbors (exact; cosine similarity) of each topic's title, description, and query variants.

To determine the threshold for the Sentence-BERT similarity score beyond which we consider a query a source of leakage for a topic, one human annotator assessed whether a query is leaking for a TREC~topic (title, description, query variants) for a stratified sample of 100~pairs of queries and topics with a similarity above~0.8. Against these manual judgments, a similarity threshold of~0.91 is the lowest that yields a 0.9~precision for deciding that a query is leaking for a topic. Table~\ref{table-candidates-leaking-queries} shows the number of topics for which queries above this threshold can be found. From MS~MARCO and ORCAS combined, 3,960~queries are leakage candidates for one of 181~Robust04 topics (72\%~of the 250~topics). From the two Common Core tracks, 37~and 38~topics have leakage candidates (76\%~of the 50~topics, respectively)---high similarities mostly against the query variants.

\begin{table}[bt]
\setlength{\tabcolsep}{.7em}
\caption{Number of topics~(T) in Robust04 and the TREC~2017 and~2018 Common Core tracks for which similar queries~(number as Q) in MS~MARCO~(MSM) and the union of MSM and ORCAS~(+ORC) exist in terms of the query having a Sentence-BERT~score~>~0.91 against the topic's title, description, or a query variant.}
\label{table-candidates-leaking-queries}
\begin{center}
\begin{tabular}{@{}l@{\quad}cc@{\quad}cc@{\quad}cc@{\quad}cc@{\quad}cc@{\quad}cc@{}}
\toprule
\textbf{Candidates} & \multicolumn{4}{@{}c@{}}{\textbf{Robust04}} & \multicolumn{4}{@{}c@{}}{\textbf{Core~2017}} & \multicolumn{4}{@{}c}{\textbf{Core~2018}} \\
\cmidrule(r{1em}){2-5} \cmidrule(r{1em}){6-9} \cmidrule{10-13}
& \multicolumn{2}{@{}c@{\quad}}{MSM} & \multicolumn{2}{@{}c@{\quad}}{+ ORC} & \multicolumn{2}{@{}c@{\quad}}{MSM} & \multicolumn{2}{@{}c@{\quad}}{+ ORC} & \multicolumn{2}{@{}c@{\quad}}{MSM} & \multicolumn{2}{@{}c}{+ ORC}\\
\cmidrule(r{1em}){2-3} \cmidrule(r{1em}){4-5} \cmidrule(r{1em}){6-7} \cmidrule(r{1em}){8-9} \cmidrule(r{1em}){10-11} \cmidrule{12-13}
& T & Q & T & Q & T & Q & T & Q & T & Q & T & Q \\
\midrule
Title       &           33 & \phantom{1}83 &           140 &           1,775 & 2 &           12 &           23 &           176 & 2 & \phantom{2}2 &           21 &           110 \\
Description & \phantom{4}2 & \phantom{11}3 & \phantom{16}8 & \phantom{3,3}50 & 0 & \phantom{1}0 & \phantom{3}0 & \phantom{60}0 & 0 & \phantom{2}0 & \phantom{2}1 & \phantom{97}2 \\
Variants    &           45 &           116 &           167 &           3,356 & 6 &           16 &           38 &           602 & 9 &           26 &           38 &           973 \\
\cmidrule[.5pt]{1-13}
Union       &           53 &           151 &           181 &           3,960 & 7 &           18 &           38 &           645 & 9 &           26 &           38 &           973 \\
\bottomrule
\end{tabular}
\end{center}
\end{table}

Some of these leakage candidates still are false positives (threshold precision of~0.9). To only use actual leaking queries in our train--test leakage experiments, two annotators manually reviewed the 5~most similar candidates per topic above the 0.91~threshold (a total of 827~candidates; not all topics had 5~candidates). Given the title, description, and narrative of a topic, the annotators labeled the similarity of a query to the topic title according to \citeauthor{jansen:2009}'s reformulation types~\cite{jansen:2009}: a query can be \emph{identical} to the topic title (differences only in inflection or word order), be a \emph{generalization} (subset of words), a \emph{specialization} (superset of words), a \emph{reformulation} (some synonymous terms), or it can be on a \emph{different topic}. An initial kappa test on 103~random of the 827~candidates showed moderate agreement (Cohen's kappa of~0.59; 103~queries: we aimed for 100~but included all queries for a topic when one was selected). After discussing the 103~cases with the annotators, they agreed on all cases and we had them each independently label half of the remaining 724~candidates. Table~\ref{table-verified-leaking-queries} shows the annotation results: 172~topics of Robust04 (i.e.,~69\%) have manually verified leaking queries (648~total), as well as 37~topics of Common Core~2017~(74\%) and 38~of Common Core~2018~(76\%). A large portion of the true-positive leaking queries are identical to or specializations of a topic's title (57.5\%~of~721). In our below train--test leakage experiments, we only use manually verified true-positive leaking queries as the source of leakage from MS~MARCO / ORCAS.

\begin{table}[bt]
\caption{Statistics of the 827~manually annotated leakage candidate queries. \Na Number of true and false candidates. \Nb Annotated query reformulation types.}
\label{table-verified-leaking-queries}
\setlength{\tabcolsep}{3pt}
\renewcommand{\arraystretch}{0.9}
\hfill%
\begin{tabular}[t]{@{}lccc@{}}
\multicolumn{4}{@{}l@{}}{\Na Manually annotated candidates.} \\
\toprule
\textbf{Corpus}            & \textbf{Candidates} & \textbf{Queries} & \textbf{Topics}\\

\midrule
\multirow{2}{*}{Robust04}  & true  &           648 &           172 \\
                           & false & \phantom{0}93 & \phantom{0}53 \\
\midrule
\multirow{2}{*}{Core 2017} & true  &           138 & \phantom{0}37 \\
                           & false & \phantom{0}21 & \phantom{0}11 \\
\midrule
\multirow{2}{*}{Core 2018} & true  &           157 & \phantom{0}38 \\
                           & false & \phantom{0}19 & \phantom{00}7 \\
\bottomrule
\end{tabular}%
\hfill%
\renewcommand{\arraystretch}{1.11}%
\begin{tabular}[t]{@{}lc@{}}
\multicolumn{2}{@{}l@{}}{\Nb Reformulation types.} \\
\toprule
\textbf{Type}   & \textbf{Queries}\\
\midrule
Identical       & 187 \\
Generalization  & 124 \\
Specialization  & 228 \\
Reformulation   & 182 \\
\midrule
Different Topic & 106 \\
\bottomrule
\end{tabular}%
\hspace*{1cm}%
\hfill

\end{table}

\section{Experimental Analysis}
\label{sec:experimental-analysis}

Focusing on zero-shot settings, we train neural retrieval models on specifically designed datasets to assess the effect of train--test leakage from MS~MARCO~/~OR\-CAS to Robust04 and TREC~2017 and~2018 Common Core. We analyze the models' effectiveness in five-fold cross-validation experiments, report detailed results for varying training set sizes for monoT5 (which has the highest effectiveness in our experiments), and study the effects of leaked instances on the retrieval scores and the resulting rankings.

\paragraph{Training Datasets.} 
For each of the three test scenarios (Robust04 and the two Common Core scenarios), we construct three types of training datasets: %
\Ni
`No Leakage' with random non-leaking queries (balanced between MS~MARCO and ORCAS as in previous experiments~\cite{craswell:2020}), %
\Nii
`MSM~Leakage' with a fixed number of random manually verified leaking queries from MS~MARCO~/~ORCAS (500~queries for Robust04, 100~queries for Common Core) supplemented by no-leakage queries till a desired size is reached, and %
\Niii
`Test Leakage' with a fixed number of queries from the actual test data (500~for Robust04, 100~for Common Core; oversampling: each topic twice (but different documents) to match `MSM Leakage') supplemented by no-leakage queries till a desired size is reached. `Test Leakage' is meant as an ``upper bound'' for any train--test leakage effect.

For each type, we construct datasets with 1,000~to 128,000~instances (500~to 64,000~queries; each with one relevant and one non-relevant document). Since MS~MARCO~/~ORCAS queries only have annotated relevant documents, we follow \citet{nogueira:2020} and sample ``non-relevant'' instances from the top-100 BM25~results for such queries. For the `Test Leakage' scenario, we use the actual TREC~judgments to sample the non-/relevant instances. In our 72~training datasets (3~test scenarios, 3~types, 8~sizes), the number of leaked instances is held constant to analyze the effect of a decreasing (and thus more realistic) ratio of leakage. Larger training data would result in more costly training, but our chosen sizes already suffice to observe a diminishing effect of train--test leakage.

\paragraph{Trained Models.}
For our experimental analyses, we use models based on mono\-BERT~\cite{nogueira:2019} and monoT5~\cite{nogueira:2020} as implemented in PyGaggle~\cite{lin:2021b}, and models based on Duet~\cite{mitra:2017}, KNRM~\cite{xiong:2017}, and PACRR~\cite{hui:2017} as implemented in Capreolus~\cite{yates:2020a} (default configurations each). In pilot experiments with 32,000~`No Leakage' instances, these models had higher nDCG@10 scores on Robust04 than CEDR~\cite{macavaney:2019}, HINT~\cite{fan:2018}, PARADE~\cite{li:2020b}, and TK~\cite{hofstaetter:2020}. Following \citet{nogueira:2020}, we use the base versions of BERT and T5 to spend the computational budget on training many models instead of a single large one. Since the training is not deterministic, each model is trained on each of the 72~training sets five times for one epoch with varying seeds (used to shuffle the training queries; configured in PyTorch). We use \query{ir\_datasets}~\cite{macavaney:2021} for data wrangling and, following previously suggested training regimes~\cite{nogueira:2020,nogueira:2019,yates:2020a}, pass the relevant and the non-relevant document of a query consecutively to a model in the same batch during training. During inference, all models re-rank the top-100 BM25 results (Capreolus, default configuration) and we break potential score ties in rankings via alphanumeric ordering by document~ID (with random~IDs, this leads to a random distribution for other document properties such as text length~\cite{lin:2019}).

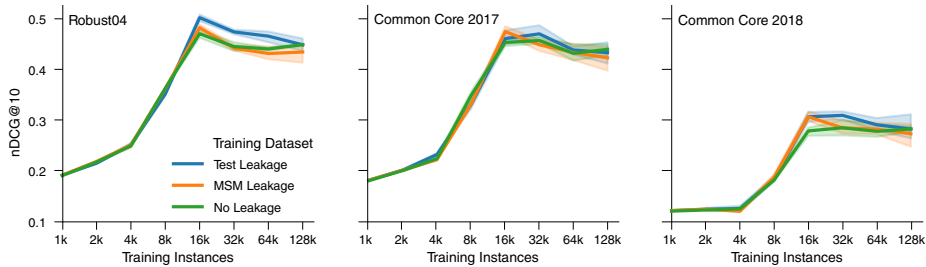
\bsfigure[width=\textwidth]{effectiveness-t5-only-leaked-topics}{Effectiveness of monoT5 measured as nDCG@10 on the topics with leakage (172~topics for Robust04, 37 and 38 for the 2017~and 2018~editions of the Common Core track). Models trained on datasets of varying size with no leakage~(No), leakage from MS~MARCO~/~ORCAS~(MSM), or leakage from the test data~(Test).}

\paragraph{Leakage-Induced nDCG~Improvements for MonoT5.}
Figure~\ref{effectiveness-t5-only-leaked-topics} shows the average nDCG@10 of monoT5, the most effective model in our experiments, for different training set sizes, tested only on topics with leaked queries. For small training sets, monoT5 achieves rather low nDCG@10 values and cannot exploit the leakage. The nDCG@10 increases with more training data on all benchmarks, peaking at~16,000 or 32,000~instances. At the peaks, monoT5 trained with leakage is more effective than without, and training on test leakage leads to a slightly higher nDCG@10 than leakage from MS~MARCO~/~ORCAS~(MSM). However, the difference between test and MSM~leakage is larger for Robust04 (with some documents published as early as~1989) compared to the newer Common Core tracks (with documents published closer to the publication date of MS~MARCO). On the Common Core data, MSM~leakage is almost as effective as test leakage.

\paragraph{Leakage-Induced Effectiveness Improvements for Other Models.}
We employ a five-fold cross-validation setup for Duet, KNRM, monoBERT, monoT5, and PACRR to study whether leakage-induced effectiveness improvements can also be observed for other models when a grid search in the cross-validation setup can choose the training set size with the highest leakage effect for each model. We report the effectiveness of the models as nDCG@10, Precision@1, and the mean first rank of a relevant document~(MFR)~\cite{fuhr:2017}.%
\footnote{We use MFR instead of the mean reciprocal rank~(MRR) as suggested by \citet{fuhr:2017}. His criticism of MRR was recently supported by further empirical evidence~\cite{zobel:2020}.}
While effectiveness scores measured via nDCG@10 and Precision@1 have the property that higher values are better (a score of~1 indicates ``best'' effectiveness), for~MFR, lower scores are better---but still a score of~1 is the best case indicating that the document on rank~1 always is relevant. In all effectiveness evaluations, we conduct significance tests via Student's t-test~($p = 0.05$) with Bonferroni correction for multiple testing as implemented in PyTerrier~\cite{macdonald:2021}.

\begin{table*}[bt]
\setlength{\tabcolsep}{2pt}
\caption{Effectiveness on Robust04~(R04) as nDCG@10, mean first rank of a relevant document~(MFR), and Precision@1 (Prec@1) in a five-fold cross-validation setup on all test topics. Models are trained with no leakage~(No), leakage from MS MARCO~/~ORCAS~(MSM), or leakage from the test data~(Test). Highest scores in bold; $\dag$ marks Bonferroni-corrected significant differences to the no-leakage baseline (Student's t-test, $p=0.05$). Model order swaps induced by MSM leakage in red.}
\label{table-effectiveness-robust04}%
\vspace*{-.2cm}%
\begin{center}
\begin{tabular}{@{}l@{\hspace{.6em}}ccc@{\hspace{.7em}}ccc@{\hspace{.7em}}ccc@{}}
\toprule
\textbf{Model} & \multicolumn{3}{@{}c@{\hspace{1em}}}{\textbf{nDCG@10 on R04}} & \multicolumn{3}{@{}c@{\hspace{1em}}}{\textbf{MFR on R04}} & \multicolumn{3}{@{}c@{}}{\textbf{Prec@1 on R04}} \\
\cmidrule(r{1em}){2-4} \cmidrule(r{1em}){5-7} \cmidrule(){8-10}
& No & MSM & Test & No & MSM & Test & No & MSM & Test \\
\midrule
Duet~\cite{mitra:2017}        & \color{red}0.201\phantom{$^{\dag}$} & 0.198\phantom{$^{\dag}$} & \textbf{0.224}$^{\dag}$         & 2.420\phantom{$^{\dag}$} & 2.682\phantom{$^{\dag}$} & \textbf{2.340}\phantom{$^{\dag}$} & {\color{red} 0.297\phantom{$^{\dag}$}} & 0.261\phantom{$^{\dag}$} & \textbf{0.304}\\
KNRM~\cite{xiong:2017}        & \color{red}0.194\phantom{$^{\dag}$} & \color{red}0.214$^{\dag}$           & \textbf{0.309}$^{\dag}$      & 2.348\phantom{$^{\dag}$} & 2.309\phantom{$^{\dag}$} & \textbf{1.976}$^{\dag}$  & {\color{red} 0.293\phantom{$^{\dag}$}} & {\color{red} 0.313\phantom{$^{\dag}$}} & \textbf{0.329}\\
monoBERT~\cite{nogueira:2019} & 0.394\phantom{$^{\dag}$} & 0.373$^{\dag}$           & \textbf{0.396}\phantom{$^{\dag}$} & 1.688\phantom{$^{\dag}$} & 1.725\phantom{$^{\dag}$} & \textbf{1.639}\phantom{$^{\dag}$} & 0.434\phantom{$^{\dag}$} & \textbf{0.454}\phantom{$^{\dag}$} & 0.414\\
monoT5~\cite{nogueira:2020}   & 0.461\phantom{$^{\dag}$} & 0.457\phantom{$^{\dag}$} & \textbf{0.478}$^{\dag}$      & 1.443\phantom{$^{\dag}$} & \textbf{1.416}\phantom{$^{\dag}$} & 1.417\phantom{$^{\dag}$}    & 0.562\phantom{$^{\dag}$} & 0.578\phantom{$^{\dag}$} & \textbf{0.590}\\
PACRR~\cite{hui:2017}         & 0.382\phantom{$^{\dag}$} & 0.364$^{\dag}$           & \textbf{0.391}\phantom{$^{\dag}$} & 1.663\phantom{$^{\dag}$} & 1.604\phantom{$^{\dag}$} & \textbf{1.579}$^{\dag}$ & 0.458\phantom{$^{\dag}$} & 0.462\phantom{$^{\dag}$} & \textbf{0.502}\\
\bottomrule
\end{tabular}
\end{center}%
\end{table*}

Table~\ref{table-effectiveness-robust04} shows the five-fold cross-validated effectiveness on Robust04 for the five models when optimizing each fold for nDCG@10, MFR, or Precision@1 in a grid search. Models trained on test leakage almost always are more effective than their no-leakage counterparts (exception: Precision@1 of monoBERT) and actual test leakage usually helps more than leakage from MS~MARCO~/~ORCAS (MSM; exceptions: MFR of monoT5 and Precision@1 of monoBERT). Overall, on Robust04, models trained with MSM~leakage are often less effective than their no-leakage counterparts (e.g., the nDCG@10 of monoBERT). A possible explanation might be the large time gap between the Robust04 document publication dates (documents published between~1987 and~2007) and the MS~MARCO data (crawled in~2018). A similar observation was made during the TREC~2021 Deep Learning track~\cite{craswell:2021}. The transition from Version~1 of MS~MARCO to Version~2 (crawled three years later) caused some models to prefer older documents since they saw old documents as positive instances and newer ones as negative instances during training. Still, MSM~leakage can lead to swaps in model ranking on Robust04. For instance, KNRM trained with MSM~leakage achieves a higher nDCG@10 and Precision@1 than Duet without leakage, while KNRM trained without leakage is less effective than Duet.

\begin{table*}[bt]
\setlength{\tabcolsep}{2pt}
\caption{Effectiveness on Common Core~2017~(CC17) as nDCG@10, mean first rank of a relevant document~(MFR), and Precision@1 (Prec@1) in a five-fold cross-validation setup on all test topics. Models are trained with no leakage~(No), leakage from MS MARCO~/~ORCAS~(MSM), or leakage from the test data~(Test). Highest scores in bold; $\dag$ marks Bonferroni-corrected significant differences to the no-leakage baseline (Student's t-test, $p=0.05$). Model order swaps induced by MSM leakage in red.}
\label{table-effectiveness-cc17}%
\vspace*{-.2cm}%
\begin{center}
\begin{tabular}{@{}l@{\hspace{.5em}}ccc@{\hspace{.7em}}ccc@{\hspace{.7em}}ccc@{}}
\toprule
\textbf{Model} & \multicolumn{3}{@{}c@{\hspace{1em}}}{\textbf{nDCG@10 on CC17}} & \multicolumn{3}{@{}c@{\hspace{1em}}}{\textbf{MFR on CC17}} & \multicolumn{3}{@{}c@{}}{\textbf{Prec@1 on CC17}} \\
\cmidrule(r{1em}){2-4} \cmidrule(r{1em}){5-7} \cmidrule(){8-10}
& No & MSM & Test & No & MSM & Test & No & MSM & Test \\
\midrule
Duet~\cite{mitra:2017}        & 0.374\phantom{$^{\dag}$} & 0.373\phantom{$^{\dag}$} & \textbf{0.376} & {\color{red} 1.620} & {\color{red} 1.512} & \textbf{1.485}\phantom{$^{\dag}$} & {\color{red} 0.500} & 0.480 & \textbf{0.540}\\
KNRM~\cite{xiong:2017}        & 0.316\phantom{$^{\dag}$} & \textbf{0.343}$^{\dag}$  & 0.330\phantom{$^{\dag}$}         & {\color{red} 1.587} & \textbf{1.512} & 1.568\phantom{$^{\dag}$} & {\color{red} 0.480} & {\color{red} \textbf{0.520}} & 0.480\\
monoBERT~\cite{nogueira:2019} & \color{red} 0.402\phantom{$^{\dag}$} & \color{red} 0.407\phantom{$^{\dag}$} & \textbf{0.419}\phantom{$^{\dag}$} & 1.625 & \textbf{1.605} & 1.634\phantom{$^{\dag}$} & \textbf{0.480} & 0.460 & 0.460\\
monoT5~\cite{nogueira:2020}   & 0.445\phantom{$^{\dag}$} & 0.464\phantom{$^{\dag}$} & \textbf{0.490}$^{\dag}$        & 1.363\phantom{$^{\dag}$} & 1.384 & \textbf{1.359}    & 0.660 & 0.620 & \textbf{0.680}\\
PACRR~\cite{hui:2017}         & \color{red} 0.406\phantom{$^{\dag}$} & 0.403\phantom{$^{\dag}$} & \textbf{0.413}\phantom{$^{\dag}$} & \textbf{1.390} & 1.515 & 1.546\phantom{$^{\dag}$} & 0.540 & 0.520 & \textbf{0.580}\\
\bottomrule
\end{tabular}
\end{center}%
\vspace*{-.2cm}%
\end{table*}

Table~\ref{table-effectiveness-cc17} shows the five-fold cross-validated effectiveness on the TREC~2017 Common Core track for the five models when optimizing each fold for nDCG@10, MFR, or Precision@1 in a grid search. In contrast to Robust04, more models improve their effectiveness when trained with MSM~leakage as the time gap between the New York Times Annotated Corpus and MS~MARCO is smaller than for Robust04. MonoT5 with actual test leakage is the most effective model for all three measures, and monoT5 trained on MSM~leakage is more effective than the no-leakage counterpart in nDCG@10 and MFR. MSM~leakage also may cause model order swaps at higher positions in the systems' nDCG@10 ordering: monoBERT with MSM~leakage would slightly overtake PACRR. Still, most of the effectiveness improvements on this dataset caused by MSM~leakage or test leakage are not significant (exception: the nDCG@10~differences for monoT5 with test leakage and KNRM with MSM~leakage to the no-leakage counterparts).

\begin{table*}[bt]
\setlength{\tabcolsep}{2pt}
\caption{Effectiveness on Common Core~2018~(CC18) as nDCG@10, mean first rank of a relevant document~(MFR), and Precision@1 (Prec@1) in a five-fold cross-validation setup on all test topics. Models are trained with no leakage~(No), leakage from MS MARCO~/~ORCAS~(MSM), or leakage from the test data~(Test). Highest scores in bold; $\dag$ marks Bonferroni-corrected significant differences to the no-leakage baseline (Student's t-test, $p=0.05$). Model order swaps induced by MSM leakage in red.}
\label{table-effectiveness-cc18}%
\vspace*{-.2cm}%
\begin{center}
\begin{tabular}{@{}l@{\hspace{.2em}}ccc@{\hspace{.6em}}ccc@{\hspace{.6em}}ccc@{}}
\toprule
\textbf{Model} & \multicolumn{3}{@{}c@{\hspace{1em}}}{\textbf{nDCG@10 on CC18}} & \multicolumn{3}{@{}c@{\hspace{1em}}}{\textbf{MFR on CC18}} & \multicolumn{3}{@{}c@{}}{\textbf{Prec@1 on CC18}} \\
\cmidrule(r{1em}){2-4} \cmidrule(r{1em}){5-7} \cmidrule(){8-10}
& No & MSM & Test & No & MSM & Test & No & MSM & Test \\
\midrule
Duet~\cite{mitra:2017}        & 0.285\phantom{$^{\dag}$} & \textbf{0.301}\phantom{$^{\dag}$} & 0.295\phantom{$^{\dag}$} & 1.993\phantom{$^{\dag}$} & \textbf{1.812}\phantom{$^{\dag}$} & 2.231\phantom{$^{\dag}$} & 0.320\phantom{$^{\dag}$} & \textbf{0.380}\phantom{$^{\dag}$} & 0.260\\
KNRM~\cite{xiong:2017}        & 0.201\phantom{$^{\dag}$} & \textbf{0.256}$^{\dag}$           & 0.238$^{\dag}$ & 3.099\phantom{$^{\dag}$} & \textbf{2.768}\phantom{$^{\dag}$} & 3.125\phantom{$^{\dag}$} & 0.100\phantom{$^{\dag}$} & \textbf{0.160}\phantom{$^{\dag}$} & 0.140\\
monoBERT~\cite{nogueira:2019} &  \color{red} 0.364\phantom{$^{\dag}$} &  \color{red} \textbf{0.380}\phantom{$^{\dag}$} & 0.366\phantom{$^{\dag}$} & 1.810\phantom{$^{\dag}$} & \textbf{1.683}\phantom{$^{\dag}$} & 1.719\phantom{$^{\dag}$} & {\color{red} 0.460\phantom{$^{\dag}$}} & {\color{red} \textbf{0.560}\phantom{$^{\dag}$}} & 0.460\\
monoT5~\cite{nogueira:2020}   & 0.417\phantom{$^{\dag}$} & \textbf{0.448}\phantom{$^{\dag}$} & 0.445\phantom{$^{\dag}$} & {\color{red} 1.577\phantom{$^{\dag}$}} & \textbf{1.503}\phantom{$^{\dag}$} & 1.512\phantom{$^{\dag}$}  & {\color{red}0.480\phantom{$^{\dag}$}} & {\color{red} \textbf{0.540}\phantom{$^{\dag}$}} & \textbf{0.540}\\
PACRR~\cite{hui:2017}         &  \color{red} 0.376\phantom{$^{\dag}$} & \textbf{0.406}\phantom{$^{\dag}$} & 0.393\phantom{$^{\dag}$} & {\color{red} 1.649\phantom{$^{\dag}$}} & {\color{red}\textbf{1.383}$^{\dag}$} & 1.485\phantom{$^{\dag}$} & {\color{red} 0.520\phantom{$^{\dag}$}} & \textbf{0.560}\phantom{$^{\dag}$} & 0.540\\
\bottomrule
\end{tabular}
\end{center}%
\vspace*{-.2cm}%
\end{table*}

Table~\ref{table-effectiveness-cc18} shows the five-fold cross-validated effectiveness on the TREC~2018 Common Core track for the five models when optimizing each fold for nDCG@10, MFR, or Precision@1 in a grid search. In contrast to Robust04 and the 2017~edition of the Common Core track, training with MSM~leakage improves the effectiveness in all cases for all three measures. While most of the leakage-induced effectiveness improvements are not statistically significant, the model order even changes on the top MFR~position, where PACRR with MSM~leakage would overtake monoT5 without leakage.

\paragraph{Discussion.}
The results in Tables~\ref{table-effectiveness-robust04}--\ref{table-effectiveness-cc18} show that leakage from MS~MARCO / ORCAS~(MSM) can have an impact on the retrieval effectiveness, even when only a small number of instances are leaked, as in our experiments. While the changes on Robust04 are rather negligible, the impact is larger for the Common Core tracks with document publication dates closer to the ones from MS~MARCO. Interestingly, MSM leakage-induced nDCG@10 improvements sometimes can lead to swaps in model ordering despite the improvements not being significant in most cases. This exemplifies that experimental effectiveness comparisons might be invalid when some models had access to leaked instances during training.

\begin{table*}[bt]
\setlength{\tabcolsep}{2pt}
\caption{Mean rank of the (leaked) relevant training documents ($\mathbf{\pm}$ standard deviation) for models trained with and without leakage from MS~MARCO~/~ORCAS~(MSM~leak\-age) or from the test data (test leakage). Ranks macro-averaged over all topics for test leakage and over all topics with leaking queries for MSM~leakage.}
\label{table-mean-first-leaked}
\centering
\begin{tabular}{@{}ll@{\hspace{.3em}}cc@{\hspace{.7em}}cc@{\hspace{.7em}}cc@{}}
\toprule
&\textbf{Model} & \multicolumn{2}{@{}c@{\hspace{1em}}}{\textbf{Robust04}} & \multicolumn{2}{@{}c@{\hspace{1em}}}{\textbf{Common~Core~17}} & \multicolumn{2}{@{}c@{}}{\textbf{Common~Core~18}} \\
\cmidrule(r{1em}){3-4} \cmidrule(r{1em}){5-6} \cmidrule(){7-8}
& & \multicolumn{1}{@{}l@{}}{With} & \multicolumn{1}{@{}l@{}}{Without} & \multicolumn{1}{@{}l@{}}{With} & \multicolumn{1}{@{}l@{}}{Without} & \multicolumn{1}{@{}l@{}}{With} & \multicolumn{1}{@{}l@{}}{Without} \\
\midrule
\multirow{5}{*}{\rotatebox[origin=c]{90}{\parbox[c]{6em}{\centering \textbf{MSM leak.}}}}    & Duet & 41.70\,{\tiny\color{gray}$\mathbf{\pm}$45.88} & 46.79\,{\tiny\color{gray}$\mathbf{\pm}$46.51} & 34.52\,{\tiny\color{gray}$\mathbf{\pm}$32.67} & 35.98\,{\tiny\color{gray}$\mathbf{\pm}$32.93} & 43.39\,{\tiny\color{gray}$\mathbf{\pm}$33.52} & 45.67\,{\tiny\color{gray}$\mathbf{\pm}$32.99}\\
 & KNRM & 82.36\,{\tiny\color{gray}$\mathbf{\pm}$31.88} & 84.74\,{\tiny\color{gray}$\mathbf{\pm}$30.15} & 43.24\,{\tiny\color{gray}$\mathbf{\pm}$31.74} & 43.68\,{\tiny\color{gray}$\mathbf{\pm}$31.50} & 53.12\,{\tiny\color{gray}$\mathbf{\pm}$32.14} & 53.45\,{\tiny\color{gray}$\mathbf{\pm}$32.14}\\
 & monoBERT & 23.08\,{\tiny\color{gray}$\mathbf{\pm}$28.71} & 23.58\,{\tiny\color{gray}$\mathbf{\pm}$28.22} & 46.97\,{\tiny\color{gray}$\mathbf{\pm}$34.95} & 47.11\,{\tiny\color{gray}$\mathbf{\pm}$35.49} & 41.79\,{\tiny\color{gray}$\mathbf{\pm}$36.16} & 42.48\,{\tiny\color{gray}$\mathbf{\pm}$36.39}\\
 & monoT5 & 20.13\,{\tiny\color{gray}$\mathbf{\pm}$26.77} & 20.15\,{\tiny\color{gray}$\mathbf{\pm}$26.64} & 35.68\,{\tiny\color{gray}$\mathbf{\pm}$31.69} & 36.46\,{\tiny\color{gray}$\mathbf{\pm}$31.88} & 29.86\,{\tiny\color{gray}$\mathbf{\pm}$28.24} & 30.31\,{\tiny\color{gray}$\mathbf{\pm}$28.27}\\
 & PACRR & 42.41\,{\tiny\color{gray}$\mathbf{\pm}$44.86} & 42.43\,{\tiny\color{gray}$\mathbf{\pm}$44.67} & 35.79\,{\tiny\color{gray}$\mathbf{\pm}$33.71} & 36.28\,{\tiny\color{gray}$\mathbf{\pm}$33.83} & 34.76\,{\tiny\color{gray}$\mathbf{\pm}$36.45} & 35.70\,{\tiny\color{gray}$\mathbf{\pm}$36.87}\\

\midrule
\multirow{5}{*}{\rotatebox[origin=c]{90}{\parbox[c]{6em}{\centering \textbf{Test leak.}}}}  & Duet & 90.04\,{\tiny\color{gray}$\mathbf{\pm}$26.98} & 90.65\,{\tiny\color{gray}$\mathbf{\pm}$26.41} & 45.78\,{\tiny\color{gray}$\mathbf{\pm}$30.03} & 46.55\,{\tiny\color{gray}$\mathbf{\pm}$30.34} & 46.31\,{\tiny\color{gray}$\mathbf{\pm}$29.85} & 46.35\,{\tiny\color{gray}$\mathbf{\pm}$30.13}\\
 & KNRM & 89.95\,{\tiny\color{gray}$\mathbf{\pm}$26.43} & 91.20\,{\tiny\color{gray}$\mathbf{\pm}$25.24} & 47.37\,{\tiny\color{gray}$\mathbf{\pm}$32.80} & 47.49\,{\tiny\color{gray}$\mathbf{\pm}$32.81} & 50.53\,{\tiny\color{gray}$\mathbf{\pm}$32.40} & 50.13\,{\tiny\color{gray}$\mathbf{\pm}$32.26}\\
 & monoBERT & 47.01\,{\tiny\color{gray}$\mathbf{\pm}$31.84} & 47.39\,{\tiny\color{gray}$\mathbf{\pm}$31.80} & 46.64\,{\tiny\color{gray}$\mathbf{\pm}$31.51} & 47.12\,{\tiny\color{gray}$\mathbf{\pm}$31.51} & 43.19\,{\tiny\color{gray}$\mathbf{\pm}$31.51} & 44.04\,{\tiny\color{gray}$\mathbf{\pm}$31.66}\\
 & monoT5 & 45.28\,{\tiny\color{gray}$\mathbf{\pm}$32.09} & 45.37\,{\tiny\color{gray}$\mathbf{\pm}$31.96} & 46.35\,{\tiny\color{gray}$\mathbf{\pm}$31.47} & 47.45\,{\tiny\color{gray}$\mathbf{\pm}$31.83} & 40.16\,{\tiny\color{gray}$\mathbf{\pm}$31.18} & 40.95\,{\tiny\color{gray}$\mathbf{\pm}$31.24}\\
 & PACRR & 80.89\,{\tiny\color{gray}$\mathbf{\pm}$34.05} & 82.60\,{\tiny\color{gray}$\mathbf{\pm}$33.07} & 53.59\,{\tiny\color{gray}$\mathbf{\pm}$31.49} & 52.91\,{\tiny\color{gray}$\mathbf{\pm}$31.26} & 52.25\,{\tiny\color{gray}$\mathbf{\pm}$32.69} & 52.28\,{\tiny\color{gray}$\mathbf{\pm}$32.32}\\

\bottomrule
\end{tabular}
\end{table*}

\paragraph{Memorization of Leaked Instances.}
To analyze whether the models memorize leaked instances, we compare the retrieval scores and resulting ranks of leaked documents in the test rankings when training includes or does not include leakage. For leaked documents not returned in the top-100 BM25~results---the models only re-rank these---, we determine a hypothetical rank by calculating the score of this document for the query and inserting the document at the corresponding rank in the to-be-re-ranked 100~documents (including breaking score-ties by document~ID). Each leaked document thus has a maximal rank of~101. 

Table~\ref{table-mean-first-leaked} shows the mean rank of relevant documents when they were included during training (with leakage) or not (without leakage). Models perfectly memorizing their positive training instances (i.e., relevant documents for test queries) would rank these documents at substantially higher positions than models that did not see the same instance during training. However, while the mean rank of leaked relevant documents improves for most cases, the improvement is mostly negligible. For instance, the mean rank of leaked relevant documents for the very effective monoT5 and monoBERT models improves only slightly compared to their no-leakage counterparts on all three corpora. But the difference increases (still rather negligibly, though) on the corpora on which leakage was more effective. In combination with the high standard deviations, one can hardly see memorization effects for the positions of leaked relevant documents in the final rankings. We thus also inspect the retrieval scores of the leaked documents.

\begin{table*}[bt]
\setlength{\tabcolsep}{4pt}
\caption{Mean retrieval score of the (leaked) relevant training documents ($\mathbf{\pm}$~standard deviation; higher scores = ``more relevant'') for models trained with / without leakage from MS~MARCO~/~ORCAS~(MSM) or the test data (Test). Scores macro-averaged over all topics for test leakage and over all topics with leaking queries for MSM leakage.}
\label{table-mean-score-leaked}
\centering
\begin{tabular}{@{}ll@{\hspace{1em}}cc@{\hspace{1em}}cc@{\hspace{1em}}cc@{}}
\toprule
&\textbf{Model} & \multicolumn{2}{@{}c@{\hspace{1em}}}{\textbf{Robust04}} & \multicolumn{2}{@{}c@{\hspace{1em}}}{\textbf{Common~Core~17}} & \multicolumn{2}{@{}c@{}}{\textbf{Common~Core~18}} \\
\cmidrule(r{1em}){3-4} \cmidrule(r{1em}){5-6} \cmidrule(){7-8}
& & \multicolumn{1}{@{}l@{}}{With} & \multicolumn{1}{@{}l@{}}{Without} & \multicolumn{1}{@{}l@{}}{With} & \multicolumn{1}{@{}l@{}}{Without} & \multicolumn{1}{@{}l@{}}{With} & \multicolumn{1}{@{}l@{}}{Without} \\
\midrule
\multirow{5}{*}{\rotatebox[origin=c]{90}{\parbox[c]{6em}{\centering \textbf{MSM leak.}}}}      & Duet & \phantom{-}0.89\,{\tiny\color{gray}$\mathbf{\pm}$1.22} & \phantom{-}0.78\,{\tiny\color{gray}$\mathbf{\pm}$1.18} & \phantom{-}0.52\,{\tiny\color{gray}$\mathbf{\pm}$1.18} & \phantom{-}0.47\,{\tiny\color{gray}$\mathbf{\pm}$1.17} & \phantom{-}0.16\,{\tiny\color{gray}$\mathbf{\pm}$0.79} & \phantom{-}0.09\,{\tiny\color{gray}$\mathbf{\pm}$0.75}\\
 & KNRM & -2.06\,{\tiny\color{gray}$\mathbf{\pm}$3.64} & -2.58\,{\tiny\color{gray}$\mathbf{\pm}$3.43} & -2.53\,{\tiny\color{gray}$\mathbf{\pm}$3.75} & -3.08\,{\tiny\color{gray}$\mathbf{\pm}$3.52} & -2.32\,{\tiny\color{gray}$\mathbf{\pm}$3.24} & -2.78\,{\tiny\color{gray}$\mathbf{\pm}$3.01}\\
 & monoBERT & -0.75\,{\tiny\color{gray}$\mathbf{\pm}$0.44} & -0.72\,{\tiny\color{gray}$\mathbf{\pm}$0.41} & -0.88\,{\tiny\color{gray}$\mathbf{\pm}$0.49} & -0.85\,{\tiny\color{gray}$\mathbf{\pm}$0.47} & -0.92\,{\tiny\color{gray}$\mathbf{\pm}$0.50} & -0.89\,{\tiny\color{gray}$\mathbf{\pm}$0.48}\\
 & monoT5 & -1.05\,{\tiny\color{gray}$\mathbf{\pm}$1.14} & -1.19\,{\tiny\color{gray}$\mathbf{\pm}$1.20} & -1.32\,{\tiny\color{gray}$\mathbf{\pm}$1.26} & -1.48\,{\tiny\color{gray}$\mathbf{\pm}$1.31} & -1.51\,{\tiny\color{gray}$\mathbf{\pm}$1.34} & -1.65\,{\tiny\color{gray}$\mathbf{\pm}$1.38}\\
 & PACRR & \phantom{-}2.59\,{\tiny\color{gray}$\mathbf{\pm}$3.25} & \phantom{-}2.29\,{\tiny\color{gray}$\mathbf{\pm}$3.11} & \phantom{-}2.78\,{\tiny\color{gray}$\mathbf{\pm}$3.35} & \phantom{-}2.46\,{\tiny\color{gray}$\mathbf{\pm}$3.20} & \phantom{-}2.25\,{\tiny\color{gray}$\mathbf{\pm}$3.08} & \phantom{-}1.95\,{\tiny\color{gray}$\mathbf{\pm}$2.96}\\

\midrule
\multirow{5}{*}{\rotatebox[origin=c]{90}{\parbox[c]{6em}{\centering \textbf{Test leak.}}}}    & Duet & \phantom{-}0.07\,{\tiny\color{gray}$\mathbf{\pm}$0.61} & -0.11\,{\tiny\color{gray}$\mathbf{\pm}$0.56} & \phantom{-}0.22\,{\tiny\color{gray}$\mathbf{\pm}$0.68} & -0.01\,{\tiny\color{gray}$\mathbf{\pm}$0.66} & \phantom{-}0.30\,{\tiny\color{gray}$\mathbf{\pm}$0.69} & \phantom{-}0.09\,{\tiny\color{gray}$\mathbf{\pm}$0.68}\\
 & KNRM & -2.71\,{\tiny\color{gray}$\mathbf{\pm}$3.79} & -3.41\,{\tiny\color{gray}$\mathbf{\pm}$3.71} & -2.78\,{\tiny\color{gray}$\mathbf{\pm}$3.65} & -3.28\,{\tiny\color{gray}$\mathbf{\pm}$3.63} & -3.08\,{\tiny\color{gray}$\mathbf{\pm}$4.14} & -3.59\,{\tiny\color{gray}$\mathbf{\pm}$4.14}\\
 & monoBERT & -0.91\,{\tiny\color{gray}$\mathbf{\pm}$0.45} & -1.04\,{\tiny\color{gray}$\mathbf{\pm}$0.53} & -0.85\,{\tiny\color{gray}$\mathbf{\pm}$0.44} & -0.90\,{\tiny\color{gray}$\mathbf{\pm}$0.48} & -0.85\,{\tiny\color{gray}$\mathbf{\pm}$0.46} & -0.92\,{\tiny\color{gray}$\mathbf{\pm}$0.49}\\
 & monoT5 & -1.70\,{\tiny\color{gray}$\mathbf{\pm}$1.24} & -2.37\,{\tiny\color{gray}$\mathbf{\pm}$1.53} & -1.47\,{\tiny\color{gray}$\mathbf{\pm}$1.09} & -1.98\,{\tiny\color{gray}$\mathbf{\pm}$1.37} & -1.52\,{\tiny\color{gray}$\mathbf{\pm}$1.24} & -2.01\,{\tiny\color{gray}$\mathbf{\pm}$1.50}\\
 & PACRR & \phantom{-}2.31\,{\tiny\color{gray}$\mathbf{\pm}$4.27} & \phantom{-}1.92\,{\tiny\color{gray}$\mathbf{\pm}$3.09} & \phantom{-}1.83\,{\tiny\color{gray}$\mathbf{\pm}$4.40} & \phantom{-}1.98\,{\tiny\color{gray}$\mathbf{\pm}$3.13} & \phantom{-}2.66\,{\tiny\color{gray}$\mathbf{\pm}$3.31} & \phantom{-}2.26\,{\tiny\color{gray}$\mathbf{\pm}$3.23}\\

\bottomrule
\end{tabular}
\end{table*}

Table~\ref{table-mean-score-leaked} shows the mean retrieval score of the relevant documents when they were included during training (with leakage) or not (without leakage). Models that memorize the leaked relevant training documents should increase their score, and we indeed observe that the retrieval score of leaked relevant documents increases in most cases compared to their no-leakage counterpart (exception: monoBERT for MSM~leakage and PACRR for test leakage from Common Core~2017). The difference between the score differences of leakage models and non-leakage models is larger for leakage from the test data than for MSM~leakage in~13 of the 15~cases (with a maximum difference for monoT5 from a test leakage difference of $0.67 = 2.37 - 1.70$ to an MSM~leakage difference of $0.14 = 1.19 - 1.05$). To investigate the ``full picture'' with respect to also negative leaked instances (i.e., non-relevant documents), we next also study the rank offsets between the positive and the negative leaked instances.

\begin{table*}[bt]
\setlength{\tabcolsep}{3pt}
\caption{Macro-averaged increase of the rank-offset between the leaked relevant and non-relevant documents ($\mathbf{\pm}$ standard deviation) for models trained on MSM~leakage ($\Delta$ on MSM) or on test leakage ($\Delta$ on Test) over the no-leakage variants.}
\label{table-memorization}
\begin{center}
\begin{tabular}{@{}l@{\hspace{.5em}}ccc@{\hspace{1em}}ccc@{}}
\toprule
\textbf{Model} & \multicolumn{3}{@{}c@{\hspace{1em}}}{\textbf{$\Delta$ on MSM}} & \multicolumn{3}{@{}c@{\hspace{1em}}}{\textbf{$\Delta$ on Test}}\\
\cmidrule(r{1em}){2-4} \cmidrule(){5-7}
& R04 & C17 & C18 & R04 & C17 & C18 \\
\midrule
Duet        & 6.378\,{\tiny\color{gray}$\mathbf{\pm}$32.15} & 3.119\,{\tiny\color{gray}$\mathbf{\pm}$19.17} & 2.647\,{\tiny\color{gray}$\mathbf{\pm}$19.23} & 0.809\,{\tiny\color{gray}$\mathbf{\pm}$17.69} & 1.430\,{\tiny\color{gray}$\mathbf{\pm}$19.33} & 1.023\,{\tiny\color{gray}$\mathbf{\pm}$20.10} \\
KNRM        & 0.640\,{\tiny\color{gray}$\mathbf{\pm}$19.22} & 0.979\,{\tiny\color{gray}$\mathbf{\pm}$15.23} & 0.398\,{\tiny\color{gray}$\mathbf{\pm}$14.55} & 1.335\,{\tiny\color{gray}$\mathbf{\pm}$11.75} & 0.012\,{\tiny\color{gray}$\mathbf{\pm}$14.92} & 0.140\,{\tiny\color{gray}$\mathbf{\pm}$15.18} \\
monoBERT & 0.692\,{\tiny\color{gray}$\mathbf{\pm}$17.97} & 0.076\,{\tiny\color{gray}$\mathbf{\pm}$17.19} & 0.369\,{\tiny\color{gray}$\mathbf{\pm}$20.04} & 3.886\,{\tiny\color{gray}$\mathbf{\pm}$20.39} & 0.980\,{\tiny\color{gray}$\mathbf{\pm}$17.44} & 3.497\,{\tiny\color{gray}$\mathbf{\pm}$25.98} \\
monoT5   & 0.443\,{\tiny\color{gray}$\mathbf{\pm}$8.60\phantom{0}} & 0.390\,{\tiny\color{gray}$\mathbf{\pm}$9.28\phantom{0}} & 0.789\,{\tiny\color{gray}$\mathbf{\pm}$9.91\phantom{0}} & 3.443\,{\tiny\color{gray}$\mathbf{\pm}$19.96} & 2.242\,{\tiny\color{gray}$\mathbf{\pm}$9.84\phantom{0}} & 1.819\,{\tiny\color{gray}$\mathbf{\pm}$10.98} \\
PACRR         & 0.043\,{\tiny\color{gray}$\mathbf{\pm}$19.30} & 0.764\,{\tiny\color{gray}$\mathbf{\pm}$10.93} & 0.452\,{\tiny\color{gray}$\mathbf{\pm}$12.38} & 1.952\,{\tiny\color{gray}$\mathbf{\pm}$17.71} & 0.271\,{\tiny\color{gray}$\mathbf{\pm}$16.96} & 0.753\,{\tiny\color{gray}$\mathbf{\pm}$14.16} \\
\bottomrule
\end{tabular}
\end{center}
\end{table*}

Table~\ref{table-memorization} shows the macro-averaged increase in the rank difference of the leaked relevant and non-relevant documents between models trained with and without leakage. The leakage increases the rank offset for all five analyzed models (e.g., 6.4~ranks for Duet on Robust04 with MSM~leakage). Interestingly, an in-depth inspection showed that most of the increased differences are caused by lowered ranks of the leaked non-relevant documents (e.g., 2~ranks lower for monoT5) while the leaked relevant documents improve their ranks only slightly (e.g., 0.3~ranks higher for monoT5).

\paragraph{Discussion.}
Overall, our results in Tables~\ref{table-mean-first-leaked}--\ref{table-memorization} indicate that memorization happens but has little impact. Larger memorization effects might be desirable in practical scenarios where a retrieval system that memorizes good results can simply present them when the same query is submitted again. However, for empirical evaluations in scientific publications or at shared tasks, (unintended) leakage memorization at a larger scale might still lead to unwanted outcomes.

\section{Conclusion}
\label{conclusion}

Our study of train--test leakage effects for neural retrieval models was inspired by the observation that 69\%~of the Robust04~topics, a dataset often used to test neural models, have very similar queries in the MS~MARCO / ORCAS datasets, that are often used to train neural models. At first glance, this overlap might seem alarming since train--test leakage is known to invalidate experimental evaluations. We thus analyzed train--test leakage effects for five neural models (Duet, KNRM, monoBERT, monoT5, and PACRR) in scenarios with different amounts of leakage. While our experiments show leakage-induced effectiveness improvements that may even lead to swaps in model ranking, our overall results are reassuring: the effects on nDCG@10 are rather small and not significant in most cases. They also become smaller the smaller (and more realistic) the amount of leakage among all training instances is. Still, even if only a few nDCG@10 differences were significant, we noticed a memorization effect: the rank offset between leaked relevant and non-relevant documents increased on all scenarios.

Train--test leakage should thus still be avoided in academic experiments but the practical consequences for real search engines might be different. The observed increased rank offset might be a highly desirable effect when presuming that queries already seen during training are submitted again after a model has been deployed to production. An interesting direction for future research is to enlarge our experiments to investigate more of the few cases where train--test leakage slightly reduced the effectiveness.

\bibliography{spire22-zero-shot-leakage-lit}

\end{document}